\documentclass[journal=jacsat,manuscript=article]{achemso}

\author{C. Arasa}
\affiliation{Gorlaeus Laboratories, Leiden Institute of Chemistry, Leiden University, P. O. Box 9502, 2300 RA Leiden, The Netherlands}
\alsoaffiliation{Leiden Observatory, Leiden University, P. O. Box 9513, 2300 RA Leiden, The Netherlands}

\author{M. C. van Hemert}
\affiliation{Gorlaeus Laboratories, Leiden Institute of Chemistry, Leiden University, P. O. Box 9502, 2300 RA Leiden, The Netherlands}
\author{E. F. van Dishoeck}
\affiliation{Leiden Observatory, Leiden University, P. O. Box 9513, 2300 RA Leiden, The Netherlands}
\author{G. J. Kroes}
\affiliation{Gorlaeus Laboratories, Leiden Institute of Chemistry, Leiden University, P. O. Box 9502, 2300 RA Leiden, The Netherlands}
\email{g.j.kroes@chem.leidenuniv.nl}
\title{Molecular Dynamics Simulations of CO$_{2}$ Formation in Interstellar Ices}

\begin{document}
\begin{abstract}
  CO$_{2}$ ice is one of the most abundant components in ice-coated interstellar ices besides H$_{2}$O and CO, but the most favorable  path to  CO$_{2}$ ice is still unclear. Molecular dynamics calculations on the ultraviolet photodissociation of different kinds of CO--H$_{2}$O  ice systems have been performed at 10~K in order to demonstrate that the reaction between CO and an OH  molecule resulting from H$_{2}$O photodissociation through the first excited state is a possible route to form CO$_{2}$ ice.  However,  our calculations,  which take into account  different ice surface models,  suggest that  there is another product with a higher  formation  probability 
((3.00$\pm$0.07)$\times$10$^{-2}$), which is the HOCO complex, whereas the formation of CO$_{2}$ has a probability of only (3.6$\pm$0.7)$\times$10$^{-4}$. The initial location of the CO is  key to obtain reaction and form CO$_{2}$: the CO needs to be located deep into the ice.
The HOCO complex becomes trapped in the cold ice surface   in the $trans$-HOCO minimum because it quickly loses its internal energy  to the surrounding ice, preventing further reaction to H + CO$_{2}$.  Several laboratory experiments have been carried out recently and they confirm that CO$_{2}$  can also be formed through other,  different routes. 
Here we compare our theoretical results with the data available  from   experiments  studying the formation of CO$_{2}$ through a similar pathway as ours, even though the initial conditions were not exactly the same. Our results also show that the HCO van der Waals complex can be formed through the interaction of CO with the H atom that is formed as a product of H$_{2}$O photodissociation. Thus, the  reaction of the H atom photofragment  following H$_{2}$O photodissociation with CO can be a possible route to form HCO ice.

\section{Introduction} \label{s:intro}

In cold and dense regions of interstellar clouds where new stars are formed several  molecules have been observed in the solid state, such as H$_{2}$O, CO, CO$_{2}$, NH$_{3}$, CH$_{4}$, and CH$_{3}$OH (among others) \cite{Tielens1991, Boogert2008}. 
Interstellar CO$_{2}$ ice has been detected for the first time in 1989 by d'Hendecourt and de Muizon \cite{dHendecourt1989},  confirmed by de Graauw $et.~al.$ \cite{deGraauw1996},  and it has been observed along many lines of sight with a constant CO$_{2}$/H$_{2}$O abundance ratio of about 0.3, with  a high column density   \cite{Pontoppidan2008}. 
The large amounts of CO$_{2}$  ice observed cannot  be explained by gas-phase reactions \cite{Hasegawa1992}, and also because of the large amount of solid CO found in ice  \cite{Gibb2004}, it is believed that CO$_{2}$ must be formed through solid phase  reactions. However, its formation path is not at all understood yet, and in the literature different routes to form CO$_{2}$ in ice have been proposed \cite{Loeffler2005, Jamieson2006, Bennett2009, Roser2001, Madzunkov2006, Watanabe2002, Watanabe2007, Oba2010, Ioppolo2011, Noble2011, Zins2011}.

Ground state CO can react with an electronically excited CO$^{*}$ molecule leading  to CO$_{2}$ and atomic C. This reaction has been studied experimentally in different laboratories \cite{Loeffler2005, Jamieson2006, Bennett2009}.  CO$_{2}$  can also be formed after  irradiating  CO ice with Ly$\alpha$ photons,  and subsequently  bombarding the ice with energetic protons of 200~keV \cite{Loeffler2005}, or by irradiating the CO ice with  5~keV electrons in order to simulate the effect of  cosmic ray particles, thereby breaking the CO bond and creating energetic atomic O \cite{Jamieson2006, Bennett2009}.

Another possible path to form  ground state CO$_{2}$ is by  the reaction between CO with atomic O($^{3}$P). Roser $et~al.$ \cite{Roser2001} and Madzunkov $et~al.$ \cite{Madzunkov2006} demonstrated that this is a possible route to form CO$_{2}$ using  temperature programmed desorption (TPD) experiments,  in one case  using thermal O atoms below 160~K  \cite{Roser2001}, and in  another case  using energetic O atoms (from 2 to 14~eV) \cite{Madzunkov2006} in order to overcome the high reaction barrier \cite{Grim1986}. However, Grim $et~al.$ \cite{Grim1986} demonstrated  that the reaction CO($^{1}\Sigma$)~+~O($^{3}$P)~$\rightarrow$~CO$_{2}$ does not take place on cold grain surfaces (10--20~K) due to its high activation energy (2970~K in the gas-phase \cite{Talbi2006}).

A third  mechanism is based on  the reaction   CO~+~OH~$\rightarrow$~CO$_{2}$~+~H, where the OH can result either from H$_{2}$O photodissociation (the photolysis mechanism) or thermal hydrogenation of oxygen species (the hydrogenation mechanism). Recent laboratory experiments show that both routes are efficient at low ice temperatures (10--20~K) \cite{Watanabe2002, Watanabe2007, Oba2010, Ioppolo2011, Noble2011, Zins2011}.
In 2002, Watanabe $et~al.$ \cite{Watanabe2002} measured the conversion rates of CO to CO$_{2}$ irradiating the ice with photons with an energy  close to Ly$\alpha$,  using  H$_{2}$O and D$_{2}$O mixed with CO ices (with a ratio of H$_{2}$O/CO~$\approx$~10) at 12~K and  a photon flux of the order of 10$^{14}$--10$^{15}$~photons~s$^{-1}$~cm$^{-2}$. After two hours irradiation time, corresponding to a photon dose of about 10$^{18}$~photons~cm$^{-2}$, most of the CO is converted to CO$_{2}$ with a rate constant of about 2.5~$\times$~10$^{-4}$~s$^{-1}$ as obtained from fits of  Fourier transform infrared spectroscopy (FTIR) measurements. 
They concluded that the rate constant is very small  and  suggested that  CO is not immediately converted to CO$_{2}$, probably because some intermediate like HOCO or DOCO is formed  first.  They did not detect HOCO, but in the end, over the long timescales of hours in the experiment,  this may lead to CO$_{2}$ and hydrogen or deuterium  atom, respectively. In these experiments, no isotope effects were observed.

In 2007,  Watanabe $et~al.$ \cite{Watanabe2007} also studied the formation of CO$_{2}$ through the hydrogenation mechanism as well as the  photolysis mechanism in two types of ices: a CO--H$_{2}$O  mixture (with a ratio of H$_{2}$O/CO~$\approx$~4), and a pure H$_{2}$O  ice layer  with a total thickness of about 30 monolayers  with one monolayer of CO on top at low ice temperatures (10--50~K). The measured photon flux was 5.9~$\times$~10$^{13}$~photons~s$^{-1}$~cm$^{-2}$ with an energy in the range of the  Lyman band and close to Ly$\alpha$. After irradiating the ices for two hours, several products were identified in the spectra, such as CO$_{2}$, which was the most abundant  product, but also HCOOH, H$_{2}$CO, CH$_{3}$OH, as well as HCO and CH$_{3}$CHO, the presence of which was inferred from small peaks.
 The CO$_{2}$ was assumed  to be formed from  the reaction of CO with OH, because CO cannot be dissociated at the UV photons energies used in these experiments.  
 The experimentalists  also fitted their data to a kinetic model in order to calculate the rate constants to form CO$_{2}$, H$_{2}$CO, and HCOOH from CO,  based on the  column densities of these photoproducts  and assuming that H$_{2}$O is constantly dissociated. According to their analysis, CO is converted to CO$_{2}$  with a rate constant of 3.3~$\pm$~0.17~$\times$~10$^{-4}$~s$^{-1}$ at 10~K, in close agreement with the value reported previously by Watanabe $et~al.$ \cite{Watanabe2002}.
However, after the bombardment of  of the ice with H atoms, HCO was the most abundant product, and also H$_{2}$CO  and CH$_{3}$OH, but no CO$_{2}$ was  detected in the infrared (IR) absorption spectrum.

Another recent laboratory study on the formation of CO$_{2}$ ice  through surface reactions  of  CO molecules  with  OH radicals formed after H$_{2}$O photodissociation was done by Oba $et~al.$ \cite{Oba2010}. In these experiments, CO molecules were introduced in the main chamber, and also H$_{2}$O  molecules that are dissociated  by a microwave induced plasma. The experiments were carried out at 10 and 20~K for two hours. During the irradiation time the ratio OH/CO was about 0.8, and the products were recorded by infrared reflection absorption spectroscopy (RAIRS). From the IR spectra some lines were assigned to the following products: CO$_{2}$, H$_{2}$CO$_{3}$, and $trans$-HOCO and $cis$-HOCO, which disappear at higher temperatures. 
The experimentalists concluded that CO$_{2}$ is formed by reaction  of the OH radicals (that have low energies)  with the CO matrix at low temperatures (below 20~K). However, it is expected that at higher temperatures the  OH  has  enough energy to migrate far away from the CO, and the reaction may not be feasible.

In  a different kind of experiments, Ioppolo $et~al.$ \cite{Ioppolo2011} deposited a mixture of CO:O$_{2}$ ice  on a substrate at low temperatures (15 and 20~K) and bombarded  the ice with a cold H atom beam,  leading  to the formation of thermal OH radicals, which after long timescales in the end reacted with CO to form CO$_{2}$. The products formed in the ice were monitored by means of RAIRS. 
Noble $et~al.$ \cite{Noble2011}  studied the formation of CO$_{2}$ ice in a nonporous H$_{2}$O ice and on an amorphous silicate surface through the reaction of CO with OH, where the OH radicals were formed after the hydrogenation of O$_{2}$ and O$_{3}$, and the formation of CO$_{2}$ in ice was demonstrated by means of  TPD experiments.
Zins $et~al.$ \cite{Zins2011} also studied the reactivity between CO and OH molecules at very low temperatures (3.5~K). The OH radicals were formed from discharged H$_{2}$O/He mixtures and they were mixed with a mixture of CO and CO$_{2}$. A Fourier transform infrared spectrometer was used in order to monitor all the products that were formed in the ice, such as CO$_{2}$, H$_{2}$CO, H$_{2}$O, and HO$_{2}$, among others. The formation of the HOCO complex was also observed when the OH radicals and the CO molecules were injected at the same time,  as an intermediate  to the formation of CO$_{2}$.

Here the photo-induced pathway is studied, building on our previous molecular dynamics (MD) simulations of the photodissociation of water molecules in crystalline and amorphous water ice \cite{Andersson2006, Andersson2008, Arasa2010} in order to better understand the formation of CO$_{2}$ through the reaction between CO and OH from a fundamental molecular physics point of view. 
We also wanted to prove that CO$_{2}$ can indeed be formed through this route even though
the system has to pass through a deep well (the HOCO well) on the way to the exit
channel barrier to reaction, so that the system could be trapped in the well if the energy of
HOCO is efficiently dissipated to the surrounding ice.
In our simulations,  the OH~+~CO reaction is based on the gas-phase potential energy surface developed by Lakin $et~al.$ \cite{LTSH} that gives rise to  low cross section   values for the gas-phase reactions \cite{McCormack, Valero2004a, Valero2004b} due to the barriers that the system has to overcome in order to eventually form CO$_{2}$ and H as is described later. However, in our simulations the OH radical  formed after H$_{2}$O photodissociation in many cases comes off  vibrationally excited, and this  might enhance the formation of CO$_{2}$ compared to  the gas-phase case.

We report results for MD simulations on the photoinduced reaction of the photofragment OH radical with CO
 in different kinds of CO--H$_{2}$O ice conditions at 10~K, where a single H$_{2}$O molecule is photodissociated by a single photon. This low  flux simulates   the flux of UV photons that irradiate ice-coated  grains deep inside interstellar clouds, which is of the order of 10$^{3}$ photons~cm$^{-2}$~s$^{-1}$ \cite{Herbst2009, Ehrenfreund2003, Garrod2006}, meaning one incident photon per month per grain, and therefore a photodissociation event will be finished by the time the next photon arrives.

The methods  employed in this study are explained in {\bf Methods}, the main results are presented in  {\bf Results and discussion}, and the final conclusions are given in {\bf Summary and Conclusion}.



\section{Methods}
\label{sec:methods}

Classical MD methods \cite{Allen1987} and analytical potentials based on pair potential interactions have been  used in order to describe the evolution of the interactions of all the molecules and fragments in a CO--H$_{2}$O ice system before and  after the absorption of an UV photon by one of the H$_{2}$O molecules from the ice.  All the specifics of the potentials and the switching functions are presented in the supporting material   \cite{Supporting}.

\subsection{Potentials}
\label{ssec:Potentials}

The total analytical potential energy surface (PES) for the CO--H$_{2}$O ice system used to describe the photodissociation of one of the water molecules  and the subsequent interaction of the fragments with CO and the other H$_{2}$O molecules can  be written as follows:

\begin{equation}\label{eq1}
V_{\rm{tot}} =V_{\rm{ice}}+V_{\rm{H_{2}O^{*}-H_{2}O}}+V_{\rm{H_{2}O^{*}-CO}}+ V_{\rm{H_{2}O^{*}}} 
\end{equation}\\
where

\begin{equation}\label{eq2}
V_{\rm{ice}} =V_{\rm{H_{2}O-H_{2}O}}+ V_{\rm{H_{2}O-CO}}
\end{equation}

The first term of the total potential (Eq.~\ref{eq1}) is given by  Eq.~\ref{eq2}, which   describes the intermolecular interactions between the H$_{2}$O molecules inside the ice excluding the  H$_{2}$O molecule that is photoexcited (Eq.~\ref{eq2}), which are  described by the TIP4P potential \cite{TIP4P}.  Eq.~\ref{eq2} also contains the intermolecular interactions between those H$_{2}$O and the  CO molecules, where  all CO and H$_{2}$O molecules are kept rigid, and the CO--CO intermolecular interactions if more than one CO molecule is considered in the system. In this work, as expressed by Eq.~\ref{eq2}  only one CO molecule is taken into account,  in order to limit  the complexity of the total system.

The  V$_{\rm{H_{2}O-CO}}$ potential is based on pair potentials between H$_{2}$O and CO molecules and consists of  repulsion, dispersion and electrostatic terms based on CCSD(T) calculations using an AVDZ basis set and applying BSSE corrections for 2500 different configurations. The charges of the H$_{2}$O molecules are the same as those used in the TIP4P potential (H:0.52$e$, O:0$e$, and the additional charge site M:-1.04$e$) \cite{TIP4P}. For the CO molecule we have used negative charges on C and on O (-0.47$e$ and -0.615$e$, respectively) and a compensating positive charge at the center of mass.

The second term of the total potential (Eq.~\ref{eq1}) contains  the intermolecular interactions of the photoexcited molecule, which is treated as fully flexible, with the rigid water molecules by means of a TIP3P-type potential \cite{Andersson2006, Andersson2008, TIP4P}. The photoexcited molecule will dissociate into H and OH. Thus, this term also takes into account the V$_{\rm{H-H_{2}O}}$,  V$_{\rm{OH-H_{2}O}}$ interactions as fully  described in our previous studies \cite{Andersson2006, Andersson2008}.

The third term of Eq.~\ref{eq1} covers the interactions between the photoexcited water molecule with CO, and the interactions between H with CO, and OH with CO. The  V$_{\rm{H_{2}O^{*}-CO}}$  potential is  taken to be  the same as for ground state H$_{2}$O interacting with CO because the lifetime of the excited water molecule is very short ($\sim$~0.2~fs \cite{Andersson2006}), leading to H and OH which will interact with CO by means of switching functions based on the OH bond distance as described elsewhere \cite{Andersson2006}.
For the V$_{\rm{H-CO}}$ pair potential we fit an H--CO non reactive potential by using  dispersion  and repulsive terms (more details in Ref.~\cite{Supporting}). We have used the well known LTSH potential \cite{LTSH}, which  is a six dimensional PES,  to describe the     V$_{\rm{OH-CO}}$ interaction.

To smoothly switch the intramolecular interaction of OH from the one valid when it interacts with H$_{2}$O to that valid when it  interacts with CO,  we have used a switching function \cite{HCl-H2O} based on the distance between the carbon atom with the oxygen atom of OH (more details can be found in the supporting material \cite{Supporting}).

The minimum energy path of the OH~+~CO reaction according to the LTSH PES is plotted in   Fig.~\ref{Figure1} \cite{LTSH}, showing  a complex minimum energy path (MEP).  
In the entrance channel  there is the OH--CO van der Waals minimum,  followed by a  small barrier ($\sim$~0.05~eV relative to the van der Waals minimum, the $trans$~HO--CO saddle point) and  the $trans$--HOCO well with an energy of about -1.3~eV relative to the gas-phase reactants. From the $trans$--HOCO minimum there are two possible paths that bring the system to the H~+~CO$_{2}$ product. 
The first path (solid line in Fig.~\ref{Figure1}) connects the $trans$--HOCO minimum  to another minimum (of -1.2~eV), the  $cis$--HOCO complex via  a $cis$--$trans$ barrier of $\sim$~0.4~eV   relative to  the $trans$--HOCO minimum. 
From the $cis$--HOCO minimum there is another saddle point ($cis$--H--OCO) with a much higher  barrier of about 1.3~eV   that connects the  $cis$--HOCO complex with the final products. 
This exit channel barrier has a height of about 0.07~eV relative to the gas-phase reactants.
The second MEP  (dashed line in Fig.~\ref{Figure1}) connects the $trans$--HOCO complex with another saddle point (HOCO--HCO$_{2}$) with a large barrier of about 1.6~eV that leads to the HCO$_{2}$ minimum, and this barrier has a height of about 0.34~eV relative to the gas-phase reactants.
Finally, the HCO$_{2}$ minimum is connected with the final products through an H--CO$_{2}$ stationary point that has an energy of 0.2~eV above the HCO$_{2}$ minimum. Moreover, since the LTSH PES gives rise to two minima: HOCO and H~+~CO$_{2}$, their interactions with the H$_{2}$O molecules must also be included in the total PES. 
The interaction between HOCO and H$_{2}$O  has been  taken simply  as the sum of the OH--H$_{2}$O  and the H$_{2}$O--CO interactions. However, we  fit a new PES to describe the V$_{\rm{H_{2}O-CO_{2}}}$ interactions,  consisting of the dispersion, repulsive and electrostatic terms based on CCSD(T)/aug-cc-pVTZ calculations  for 174 different configurations   \cite{Supporting}. The H--H$_{2}$O interactions were already described before \cite{Andersson2006, Andersson2008}. 

The last term of Eq.~\ref{eq1} describes the intramolecular interactions in  the water molecule that is photoexited to the first excited electronic state of gas-phase H$_{2}$O ($\rm{\tilde{A}^{1}B_{1}}$) by means of the Dobbyn and Knowles PES \cite{DK1, DK2, DK3}, which  is repulsive and  leads to  dissociation into H and OH, as  has already been explained elsewhere \cite{Andersson2006, Andersson2008}.


\begin{figure}[t]
\begin{center}
\includegraphics[width=12cm]{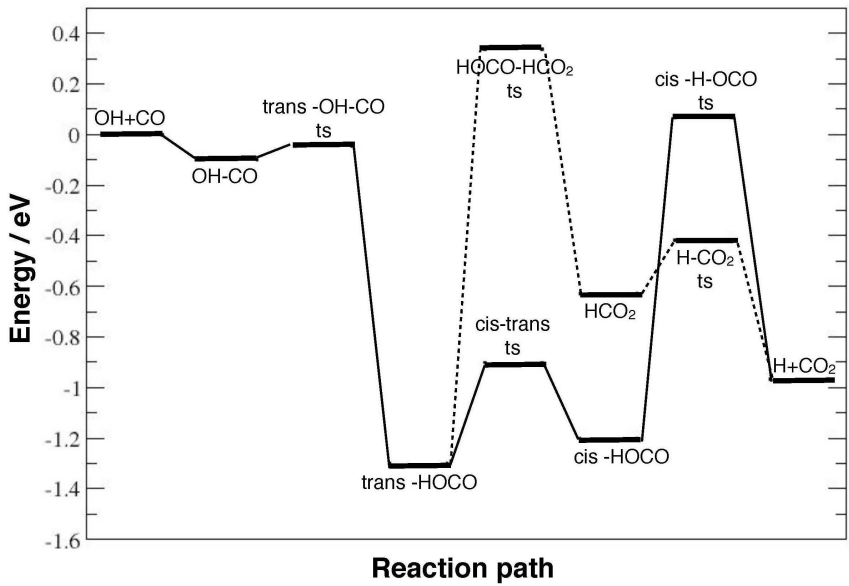}
\end{center}
\caption {{ Minimum energy path according to the LTSH potential \cite{LTSH} for the reaction OH~+~CO~$\leftrightarrow$~H~+~CO$_{2}$.  }}
\label{Figure1}
\end{figure}  

\subsection{Amorphous ice surface}
\label{ssec:Amorphous}

To study the UV photodissociation of H$_{2}$O ice in an amorphous CO--H$_{2}$O ice system followed by the reaction of OH + CO, we have used two different procedures to construct an amorphous CO--H$_{2}$O ice system.

The first procedure of growing an ice surface is based on an `hit and stick'  method. First,  a CO molecule is set up with center of mass coordinates fixed at the origin  ($x$, $y$,  $z$=0). A  water molecule is generated with random position and orientation and it interacts with the CO molecule through the H$_{2}$O--CO PES described above. The energy of the   H$_{2}$O--CO  system is minimized  by using the simplex algorithm \cite{simplex}. The CO--H$_{2}$O ice is grown
 by  a consecutive addition of  single  H$_{2}$O molecules  and minimizing the interaction energy which is described by the TIP4P PES \cite{TIP4P} for  the newly arriving H$_{2}$O with the H$_{2}$O molecules already present in the ice, and  by the H$_{2}$O--CO PES,  using the  simplex algorithm. 
After the amorphous ice is grown it is
 first thermalized, and next equilibrated at 10~K for 30~ps using MD with a thermostat switched on and off  \cite{thermostat}.  The resulting ice physically looks like an ice ball made up of one CO and  50 H$_{2}$O molecules (Fig.~\ref{Figure3}), and the closest   H$_{2}$O molecule and the farthest one from the central CO molecule are located at a distance   of 3.2~\AA, and 7.2~\AA, respectively. 
 With this methodology, we have set up three  different initial configurations of all the molecules in the ice ball, all of them thermalized at 10~K. 
The ice balls were used to explore the reactivity in the ice since they require little computer time.

The second procedure is based on our previous amorphous water ice set up  \cite{Andersson2006, Andersson2008, Arasa2010}.
Thus, first an hexagonal crystalline H$_{2}$O ice surface was modeled, and from its geometry an amorphous H$_{2}$O ice surface was obtained at 10~K  using MD simulations \cite{Allen1987} and the fast-quenching method \cite{AlHalabi2004a, AlHalabi2004b, Andersson2006}. 
The resulting ice can be divided in 16 monolayers (where the four bottom layers were not allowed to move during the dynamics in order to simulate the bulk), and in total the system contains 480~H$_{2}$O molecules with the system described by  cell parameters of 22.4~\AA, and 23.5~\AA~in $x$ and $y$, respectively \cite{Andersson2006, Andersson2008, Arasa2010}. One water molecule was replaced by a CO molecule and its random coordinates and velocities were sampled according to a Maxwell-Boltzmann distribution of 10~K. We considered three different initial scenarios for the CO molecule:  it was adsorbed on top of the first monolayer, it was absorbed in the second monolayer of the amorphous water ice surface, or it was absorbed in the fifth monolayer of the amorphous water ice surface. In the three cases, the total system was thermalized at 10~K for 20~ps using a thermostat \cite{thermostat}, and later on it was equilibrated for 20~ps  using microcanonical ensemble (NVE) MD simulations  where the CO--H$_{2}$O  and H$_{2}$O--H$_{2}$O interactions were taken into account. In Fig.~\ref{Figure2}(a),   the amorphous CO$_{ad}$--H$_{2}$O ice surface and in Fig.~\ref{Figure2}(b), and (c), the amorphous CO$_{ab}$(ML2)--H$_{2}$O ice surface and CO$_{ab}$(ML5)--H$_{2}$O ice surface are represented, respectively.

%

\begin{figure}[t]
\begin{center}
\includegraphics[width=7cm]{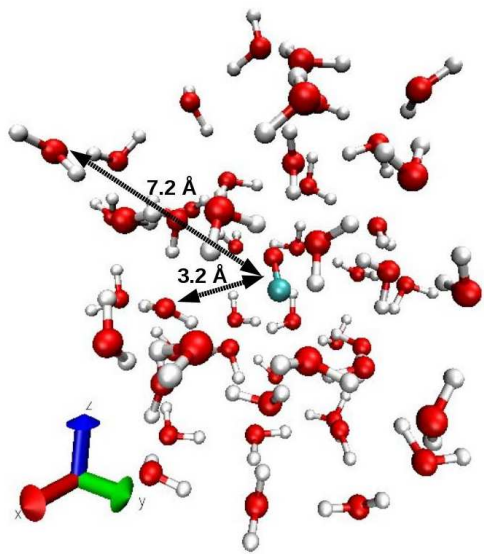}
\end{center}
\caption {{ Amorphous ice ball with 50 H$_{2}$O molecules and 1 CO molecule in the center. Red balls represent O, white H, and green C atoms.}}
\label{Figure3}
\end{figure}  
%
\newpage
\begin{figure}[htb!]
\begin{center}
\includegraphics[width=8cm]{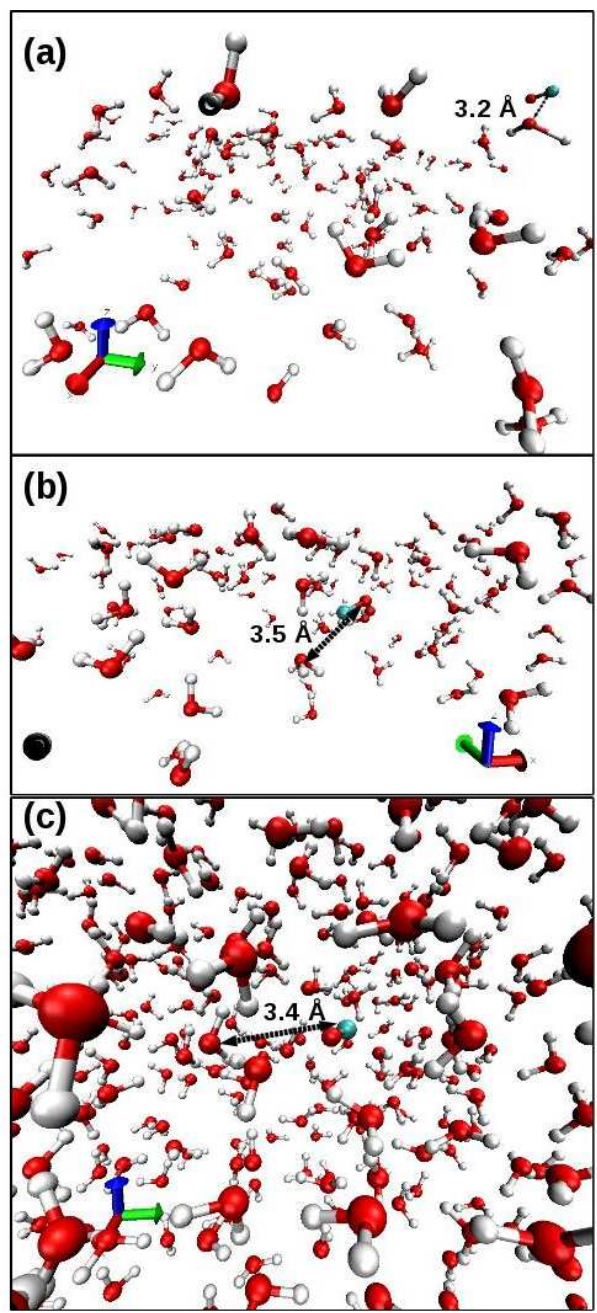}
\end{center}
\caption {{(a) Amorphous ice showing the top four MLs of H$_{2}$O  and one CO molecule adsorbed on top of ML~1. (b) Amorphous ice showing the top four MLs of H$_{2}$O  and one CO molecule absorbed in ML~2. (c) Amorphous ice showing twelve MLs of H$_{2}$O  and one CO molecule absorbed in ML~5. Red balls represent O, white H, and green C atoms.}}
\label{Figure2}
\end{figure}  

\subsection{Initial conditions and dynamics} 
\label{ssec:Initial}

For both types of CO--H$_{2}$O ice surfaces, only the  H$_{2}$O molecules that are closest  to the CO molecule (the distance between the center of mass of CO and the center of mass of water  being  smaller than 4.5~\AA)  were selected to be photodissociated, and between 200--2000 different initial coordinates and momenta were generated  per excited H$_{2}$O molecule. 
As  described elsewhere \cite{Andersson2006, Andersson2008, Arasa2010},  in order to initialize the trajectories \cite{Wigner2} a Wigner phase-space distribution function \cite{Wigner1} fitted to the ground-state vibrational wavefunction  of gas-phase water is used. The resulting coordinates and momenta were sampled  using a Monte Carlo algorithm, and  a vertical excitation is performed to put the system  on the first electronically excited state, on the DK $\rm{\tilde{A}^{1}B_{1}}$ PES  \cite{DK1, DK2, DK3}. 
Thus, the molecules are excited with   photon energies $E_{\rm{exc}}$  in  the range  7.5--9.5~eV,  with a peak at 8.6~eV (see figure~3 in Ref.~\cite{Andersson2006}). From the excitation energy and the dissociation energy of H$_{2}$O ($E_{\rm{diss}}\rm{(H_{2}O)}$ $\approx$~5.4~eV \cite{DK3}), we can estimate the initial energy of the photofragments, using that  
in the gas-phase (in the absence of the surrounding ice),  the water photofragments (H and OH) have to obey momentum conservation ($p_{\rm{H}}$=$-p_{\rm{OH}}$) and energy conservation ($E_{\rm{H}}$+$E_{\rm{OH}}$=$E_{\rm{H_{2}O}}$=$E$).   The initial available energy  $E$ can be calculated from  $E$=$E_{\rm{exc}}$$-$$E_{\rm{diss}}\rm{(H_{2}O)}$, which is in the range of 2.1--4.1~eV depending on the initial excitation energy.  The initial energy that the fragments will have can then be estimated  according to  the following equation:

\begin{equation}\label{eq3}
\frac{1}{2} m_{\rm{OH}} v^{2}_{\rm{OH}}=\frac{E}{(1 + \frac{m_{\rm{OH}}}{m_{\rm{H}}})}
\end{equation}\\
Therefore, OH radicals will be formed with approximately a maximum translational energy of  E$\rm_{OH}$=$E$/18=0.2~eV, and approximately a maximum of vibrational energy E$_{\rm{v}}$ of 2~eV  \cite{Andersson2008}.  If  the OH is formed with higher vibrational energy, the available total translational energy of OH is reduced by an `equal' amount as follows:

\begin{equation}\label{EcolOHv}
E\rm_{OH}(\upsilon)=E\rm_{OH}(\upsilon=0) - \frac{E_{\upsilon} - E_{\upsilon=0}}{18}
\end{equation}

During the photodissociation dynamics the CO molecule is treated in the same way as the water molecules that are not photodissociated, i.e.,  as a rigid rotor. 
Because the R$_{\rm{C-O'}}$ bond is always fixed to the equilibrium distance of CO the system cannot follow the actual minimum energy path to H + CO$_{2}$. However, the heights of the barriers change by less than 0.05 eV by fixing R$_{\rm{C-O'}}$, so that this does not represent a too severe approximation.  In line with this, due to its formation process OH may be vibrationally excited initially, and vibrational excitation of OH enhances the gas-phase OH + CO reaction more than vibrational excitation of CO \cite{Valero2004b, Liu2012, Li2012b}.
The initial momenta of CO are initialized from a Maxwell Boltzmann distribution of 10~K, and the whole system is equilibrated at the same temperature of 10~K.

For each trajectory only one H$_{2}$O molecule is photodissociated and Newtons's equations of motion are integrated using  a time step of 0.02~fs and a maximum time t$_{\rm{max}}$ of 5~ps.    
The simulation stops if the CO$_{2}$ molecule is formed, otherwise all the trajectories are run until t$_{\rm{max}}$.

The criterion to define the formation of the CO$_{2}$ molecule after the interaction with OH (OH + CO'~$\rightarrow$ CO$_{2}$ + H) is based on the intramolecular distances of HO--CO': $R_{\rm{C-O}}$ $\leq$~1.3~\AA, $R_{\rm{C-O'}}$ $\leq$~1.3~\AA, $R_{\rm{O-O'}}$ $\leq$~3~\AA, and the hydrogen atom originally forming OH is at a distance $R_{\rm{O-H}}$ $>$~2~\AA.
The HOCO' complex is defined to be formed if the intramolecular distances are: $R_{\rm{H-C}}$ $\leq$~2.3~\AA, $R_{\rm{H-O'}}$ $\leq$~3.3~\AA, $R_{\rm{C-O}}$ $\leq$~1.5~\AA, $R_{\rm{C-O'}}$ $\leq$~1.3~\AA, and $R_{\rm{O-O'}}$ $\leq$~3~\AA.


\section{Results and discussion}
\label{sec:results}

\subsection{Amorphous ice ball} 
\label{ssec:Ball}

We considered three different ice balls, and for each of  the fourteen closest H$_{2}$O molecules to  CO between 1000 and 2000 different initial configurations were sampled for each H$_{2}$O molecule in the  three  ice balls taking into account  the excitation energies within the first UV absorption band of amorphous water ice \cite{Kobayashi1983}.

In our study, three possible outcomes have been observed. We calculated the formation probabilities and its standard errors ($\epsilon$=$\sqrt{P\cdot(1-P)/N}$, where P is the probability and N is the total number of trajectories) for these channels  for the three different  ice balls, which lead to a total number of trajectories of $\approx$ 48,000.
The  predominant outcome channel is the non reactive one with a probability of 0.970 $\pm$ 0.001, and the second one  the formation of the HOCO complex which stays trapped in the ice,  with a probability of (2.98 $\pm$ 1.07)$\times$10$^{-2}$. The least  probable channel is   the formation of  CO$_{2}$  and a hydrogen  atom that was always observed to desorb from the ice,  with a  probability of (3.6 $\pm$ 0.9)$\times$10$^{-4}$. 
The HCO van der Waals complex formation is also observed with an energy of about -18~meV, and  a probability of (2.67 $\pm$ 0.07)$\times$10$^{-2}$,  and this complex always dissociates to CO and  H, which desorbs  while CO remains  trapped inside the ice.
The HCO van der Waals complex is defined to be formed if the distance between the H atom formed after H$_{2}$O dissociation and the carbon atom from CO $R_{\rm{H-C}}$ $\leq$~2.1~\AA, and  $R_{\rm{C-O'}}$   is always  $<$~1.3~\AA~because it is frozen.

\subsection{CO adsorbed and absorbed in amorphous water ice} 
\label{ssec:AWI}

We have also studied the formation of CO$_{2}$ and HOCO using a set up of amorphous ice
similar to the one we have used in earlier MD studies \cite{Andersson2006, Andersson2008, Arasa2010} of desorption of
H, OH and H$_{2}$O from ice following photodissociation of an H$_{2}$O molecule in ice.
 In this Section we discuss an amorphous water ice surface set up at 10~K and a CO molecule that is either adsorbed on top of the first monolayer or located in the second or fifth monolayer (as we described above in the Methods Section).  The  closest H$_{2}$O molecules 
 to the CO molecule were selected to be photoexcited and between 200--2000 different initial conditions were sampled per excited H$_{2}$O molecule. 

If the CO molecule was initially adsorbed on top of the first monolayer of an amorphous water ice (Fig.~\ref{Figure2}(a)) no reaction was observed. The final outcome of the  1000 trajectories run in total was  that CO stays adsorbed  on top of the water ice surface, whereas the H atom  always desorbs and the OH fragment either desorbs or stays trapped in the ice, as in our pure water photodissociation results \cite{Andersson2006, Andersson2008, Arasa2010}.   From the previous calculations on the ice ball system, we conclude that no more than 1000  trajectories were necessary in order to observe some reaction (HOCO that is a prerequisite for the CO$_{2}$ formation).

If the CO molecule was initially located in the second monolayer of the amorphous water ice surface (Fig.~\ref{Figure2}(b)), 2000 trajectories were run in total and  only one special event was observed to take  place: the HCO van der Waals complex is formed \cite{Supporting} --even though we have used a non reactive HCO PES-- for a few femtoseconds ($\sim$~10~fs) after which the complex dissociates  to H atom and CO molecule both absorbed in the ice at different locations, and  the OH fragment remains trapped in the ice. This channel occurs with a probability of (3.15 $\pm$ 0.39)$\times$10$^{-2}$. Watanabe $et~al.$ \cite{Watanabe2007} also observed HCO experimentally after irradiation of a mixed H$_{2}$O--CO ice  with UV photons in the range of the Lyman band only for short irradiation times ($\leq$~10~min). Therefore, the photolysis of a mixed CO--H$_{2}$O ice seems to be another possible surface reaction route to  HCO formation in interstellar ices. This can be investigated in the  future with the use of a reactive H--CO potential.

If the CO molecule was initially located in the fifth monolayer of the amorphous water ice surface (Fig.~\ref{Figure2}(c)) three different outcomes were observed after running a total of 20,000 trajectories:  the non reactive one with a probability of 0.970 $\pm$ 0.001,  the formation of the HOCO complex with a probability of (2.97 $\pm$ 0.12)$\times$10$^{-2}$, and also the formation of the CO$_{2}$ molecule while the H atom desorbs from the ice surface with a probability of (3.5 $\pm$ 1.3)$\times$10$^{-4}$. Also in this case  the transient  formation of the HCO van der Waals complex has been observed,  with a probability of (2.81 $\pm$ 0.12)$\times$10$^{-2}$.

 Therefore, these findings suggest that the initial location of the CO is very important to its subsequent chemistry. 
If the CO is at the surface or near the surface, a nearby OH photofragment formed upon H$_{2}$O photodissociation will usually be in the top two monolayers of the ice surface, and we know from previous MD simulations \cite{Andersson2006, Andersson2008, Arasa2010}  that these OH fragments tend to desorb to the gas-phase rather than stay trapped in the ice. However, if the OH fragment formed upon H$_{2}$O photodissociation was initially located more deeply in the ice it will stay trapped and it is more probable that it will find the right orientation towards the CO molecule to react. The CO molecule and the OH fragment must be located deeply in the water ice system to enable HOCO or CO$_{2}$ formation from the reaction with an OH photofragment.

\subsection{HOCO and CO$_{2}$ formation} 
\label{ssec:formation}

The most important photoproduct formed in the ice after the reaction of the OH radical with the CO molecule for both types of ice  is the HOCO complex, which loses its energy during the dynamics by dissipation  to the surrounding H$_{2}$O molecules. In these cases, the  HOCO complex is trapped in the $trans$-HOCO potential well (see Fig.~\ref{Figure1}). After enough energy has been transferred to the surrounding ice at 10~K, it becomes  nearly  impossible to overcome the barriers and form CO$_{2}$ and hydrogen, as  demonstrated in Fig.~\ref{Figure4}, where we have plotted the total HOCO energy as a function of the time for one  particular trajectory.   Fig.~\ref{Figure4} shows   how fast (in less than 0.2~ps) the HOCO complex may lose its  energy,  after which it became trapped  in the potential well with a total  energy of about -1~eV in the case shown (which is close to the $trans$-HOCO minimum energy value, according to the LTSH PES \cite{LTSH}, see also  Fig.~\ref{Figure1}).

%

\begin{figure}[t]
\begin{center}
\includegraphics[width=10cm]{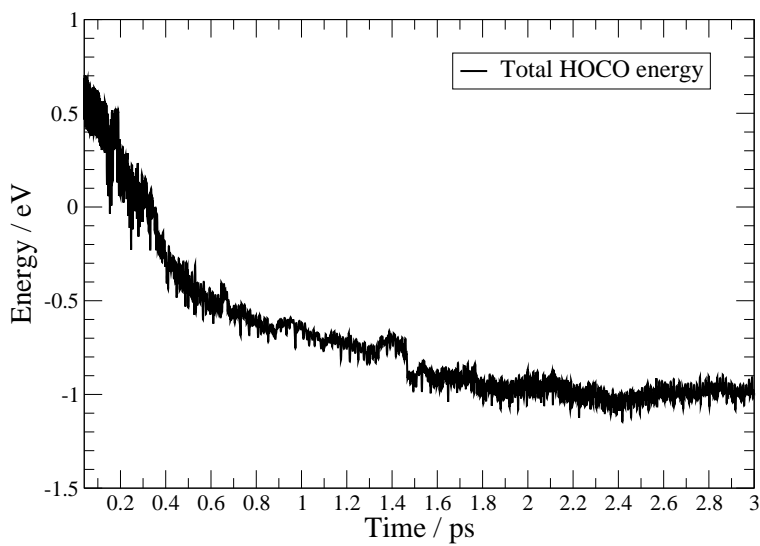}
\end{center}
\caption {{Evolution of the total HOCO energy as a function of time in ps. }}
\label{Figure4}
\end{figure}  

Goumans $et~al.$ \cite{Goumans2008} already proposed that the reaction between OH and CO may  not lead to H~+~CO$_{2}$ when the HOCO complex is formed, because its energy may be dissipated to the surrounding H$_{2}$O molecules, after which HOCO can react with atomic H to form CO$_{2}$ + H$_{2}$, or H$_{2}$O and CO, or HCOOH. 

The formation of the HOCO complex has been  observed by Oba $et~al.$ \cite{Oba2010} in their absorption spectra in both the $trans$ and $cis$ configurations through peaks at 1812~cm$^{-1}$ and 1774 ~cm$^{-1}$, respectively. These peaks were   observed earlier at 1833~cm$^{-1}$ and 1797~cm$^{-1}$ by Milligan and  Jacox  \cite{Milligan1971} during vacuum ultraviolet photolysis  of H$_{2}$O in a CO matrix  in the 200--300~nm range.
Moreover, Watanabe $et~al.$ \cite{Watanabe2002} assumed that the intermediate was formed even though they did not detect it, in order to explain the slow formation rate of CO$_{2}$ ice from CO and OH.
In another experiment, Noble $et~al.$ \cite{Noble2011} measured the yield of $^{13}$CO$_{2}$ to be only about 8~$\%$ with respect to $^{13}$CO. 
This low yield could be explained   assuming that  the HOCO intermediate is formed  as predicted  by 
gas-phase theoretical   \cite{Francisco2010, Li2012, Li2012b, Song2006, LTSH, Yu2001}  and   experimental studies \cite{Alagia1993}, through the reaction of CO with OH.  
In an ice environment, the reactivity becomes even more complicated than in the gas-phase because it also depends on the initial orientation of the H$_{2}$O  and CO reactants inside the ice before H$_{2}$O is photolysed, and on the resulting orientation of the OH photofragment relative to CO.
Francisco $et~al.$ \cite{Francisco2010} studied the stability of the HOCO complex in the gas-phase, and they found  that HOCO is stable, in agreement with the far infrared laser magnetic resonance experiment \cite{Sears1993} that confirms that HOCO is stable with a lifetime of  about 10~ms. The HOCO complex has two conformers:  $trans$-HOCO and  $cis$-HOCO.  The first one is about 0.1~eV more stable than the second one (Fig.~\ref{Figure1}). The geometry of both conformers is rather similar: one of the C--O bonds is about 1.34--1.35~\AA~long,  the other one is shorter at 1.18--1.19~\AA, and the O--H distance is about  0.97~\AA, with 107$^{\circ}$--108$^{\circ}$, and 127$^{\circ}$--130$ ^{\circ}$ HOC, and OCO bond angles, respectively \cite{Francisco2010, Li2012, Li2012b}.

According to our simulations and taking into account the two types of ice,  the probability to form the HOCO complex  is (3.00 $\pm$ 0.07)$\times$10$^{-2}$. On average the HOCO complex that is observed trapped in the ices has the following geometry: one of the C--O bonds is about 1.37~\AA~long,  and the other  CO distance corresponding to the original  CO molecule is always kept fixed, because in our simulations the CO is treated as a rigid rotor, at a bond length  of  1.13~\AA.  The O--H distance is about  0.91~\AA, the HOC angle is about 120$^{\circ}$--136$^{\circ}$,  this range being somewhat larger than in the gas-phase,   the OCO bond angle is about 123$^{\circ}$ on average,  which is close to the gas-phase value, and the dihedral HO--CO angle is on average 175$^{\circ}$. Therefore, the HOCO complex that is observed as trapped in the ice matrix has a geometry close to the $trans$-HOCO complex \cite{Li2012, Li2012b}, and its energy (see Fig.~\ref{Figure4}) agrees with the minimum of the $trans$-HOCO complex in  the LTSH potential \cite{LTSH}  used (see Fig.~\ref{Figure1}). The fixed CO bond distance in our simulations does not affect the minimum reaction barrier height relative to the gas-phase reactants (i.e., with this approximation the exit channel barrier height is only 0.02~eV higher).

The total calculated probability to form CO$_{2}$ + H for the two kinds of ice investigated is (3.6 $\pm$ 0.7)$\times$10$^{-4}$, which is very low and the reaction only occurs when the initial excitation energy of the photoexcited water molecule  is on average (8.83 $\pm$ 0.23)~eV.
At these energies usually vibrationally excited OH is produced 
 on average at $v$=1 with a vibrational energy   E$_{\rm{v}}$~$\approx$~0.7~eV.
We analyzed the initial configurations and orientations of the  reactant molecules that lead to CO$_{2}$ and the main conclusions  are that the initial distance between the C atom from CO and the oxygen atom from the reacting OH fragment has to be within the range  3.1~\AA~$\le$ R$_{\rm.{CO}}$ $\le$ 3.5~\AA, and the dihedral angle between the OH bond and CO bond has to be in the range 41$^{\circ}$ -- 178$^{\circ}$ as is illustrated in Fig.~\ref{Figure6} where the angle is 94.5$^{\circ}$.
During the dynamics OH approaches  CO and  orientates itself in a $trans$ conformation relative to CO, and in less than 0.05~ps and without going through the $cis$-HOCO conformation the OH  bond breaks and  CO$_{2}$ and H are formed.

\begin{figure}[t]
\begin{center}
\includegraphics[width=8cm]{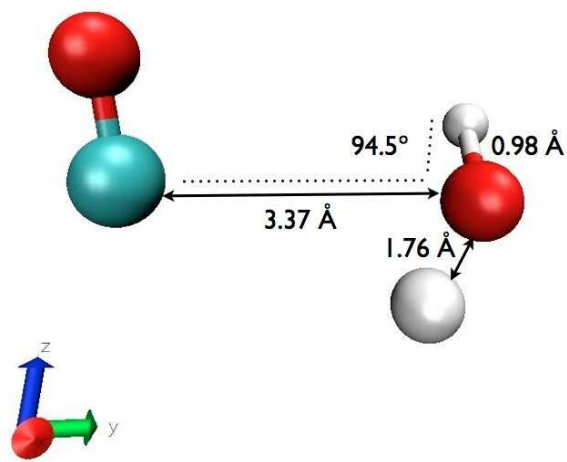}
\end{center}
\caption {{ Initial configuration of CO and OH after H$_{2}$O photodissociation that leads to the formation of CO$_{2}$. Red balls represent O, white H, and green C atoms.}}
\label{Figure6}
\end{figure}  


We compared the probability to form CO$_{2}$ in the ice with  gas-phase calculations  performed using the VENUS96 code \cite{venus96} and using the LTSH PES \cite{LTSH} implemented as described in Ref.~\cite{Valero2004a}. The  collision energy of OH was set to 0.17~eV,  which corresponds to the initial energy   OH fragments would have on average in the ice if the water molecules were initially excited  with an energy of 8.6~eV  (corresponding  to the peak of the first UV absorption band of amorphous water ice \cite{Kobayashi1983}),  and the impact parameter range was  between  0  and  2.4~\AA. 

Fig.~\ref{Figure5} presents the fitted reaction probability as a function of impact parameter for four values of $\upsilon$. For each value of $\upsilon$, the initial available total translational energy of OH is calculated according  to   Eq.~\ref{EcolOHv} (e.g., for $\upsilon$=0, 1, 2, and 3, E$\rm_{OH}$($\upsilon$)=0.17, 0.15, 0.13,  and 0.11 eV, respectively).
From Fig.~\ref{Figure5} it is clear that increasing the vibrational
energy of the reactants increases the reaction probability as well even though the translational energy of OH is reduced, in agreement with  very recent  theoretical gas-phase calculations \cite{Liu2012} where a full-dimensional time-dependent wave packet study based on the LTSH potential \cite{LTSH}  showed how the reactivity is enhanced by initial OH vibrational excitation.  
The probabilities to form CO$_{2}$ in the gas-phase for OH in $\upsilon$=0, 1,  2, and 3 are 0.01, 0.037,   0.063, and 0.082, respectively
 at $b$=0~\AA~(Fig.~\ref{Figure5}). 
The most reactive trajectories are those  that start with the OH radical more or less in line with the CO molecule, O of OH closes to C of CO.
 Therefore,  in the gas-phase the probability to form CO$_{2}$ and hydrogen from  OH + CO is higher than in our calculations. But   these probabilities are for impact parameter  $b$=0~\AA, and they decrease when the impact parameter increases up to a maximum impact parameter of about 2~\AA~(Fig.~\ref{Figure5}), whereas in our simulations when water dissociates in ice into OH and H, the OH radical collides with CO at different impact parameters.
In particular, the calculated impact parameter $b$ for the reactive trajectories was within the range  1.27--4.4~\AA, on average being  (2.94 $\pm$ 0.86)~\AA. 
Moreover,  in the solid state the environment is also completely different than in the gas-phase because the OH and the CO molecules also interact with the other H$_{2}$O molecules in the ice and they may release their energy to the environment, which may also   decrease the   reactivity.


\begin{figure}[t]
\begin{center}
\includegraphics[width=12cm]{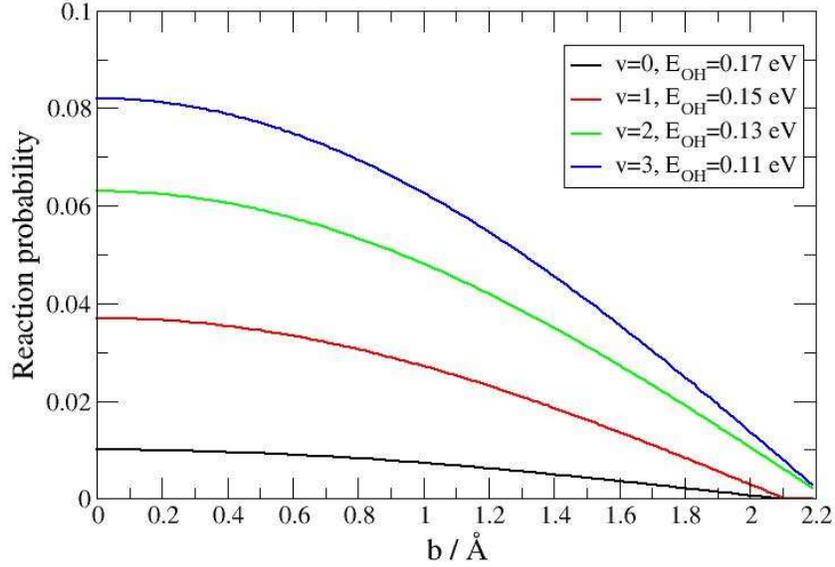}
\end{center}
\caption {{Gas-phase fitted reaction probability on the LTSH PES \cite{LTSH} versus the  impact parameter at $\upsilon$=0, 1, 2, and 3. }}
\label{Figure5}
\end{figure} 

We have analyzed  our simulations for the two types of ice systems and  found that initially the OH vibrational state  distributions mostly correspond to  $\upsilon$=0, 1, 2, and 3, using box  quantization to assign vibrational states with inclusion of zero-point energy. The OH  vibrational energy distribution  agrees  with the experimental gas-phase distribution  \cite{Hwang1999}  for  H$_{2}$O photodissociation  at 7.9~eV (this value was chosen because the production of OH in ice is shifted by the same amount as the absorption spectrum). The experimental data show that when H$_{2}$O dissociates several vibrational states are assigned to the OH photoproduct with  $\upsilon$=0, 1, 2, 3, and 4.

\subsection{Comparison with the experiments} 
\label{ssec:Exp}

The comparison with the experiments that study the formation of CO$_{2}$ ice through the reaction of CO with OH, where the OH comes from UV dissociation of H$_{2}$O ice \cite{Watanabe2002, Watanabe2007},  is not straightforward. The experiments only provide the rates to convert CO into CO$_{2}$ based on the variation of the column densities as a function of irradiation time, and from these data a first-order  rate  equation law is fit. 
On the other hand, from our theoretical models we can  provide  the formation probabilities of CO$_{2}$  and HOCO in ice per absorbed UV photon, which are 
(3.6 $\pm$ 0.7)$\times$10$^{-4}$, and (3.00 $\pm$ 0.07)$\times$10$^{-2}$  (Table~\ref{table1}), respectively, where it is assumed that,  as in the gas-phase photoabsorption, each absorbed photon leads to dissociation into H and OH.

In order to  estimate the same probability from the experiments performed by Watanabe $et~al.$ \cite{Watanabe2002}, we need,  besides the photon flux used in the experiments, the number of water and CO molecules in the sample, and the photon absorption cross section ($\sigma$) of amorphous ice  in the Ly$\alpha$ (10.21~eV) region.
Unfortunately, $\sigma$ is not accurately known. Ref.~\cite{sigma1} claims  it to be close to the absorption coefficient of liquid water and around 30$\%$ of  the gas-phase value. Accurate values of the gas-phase cross section are given in Ref.~\cite{sigma2}  at exactly Ly$\alpha$ and in Ref.~\cite{Cheng2004} for the whole VUV range.
Here we use a gas-phase cross section, $\sigma_{\rm{gas}}$, averaged over the 116--136~nm range used in the experiment \cite{Cheng2004},  $\sigma_{\rm{gas}}$=7.1$\times$10$^{-18}$~cm$^{2}$. 
Let $I\rm{_{0}}$  be the experimental photon flux per cm$^{2}$ per second. Then the number of photons absorbed per second per cm$^{2}$ follows from Beer's law as:

\begin{equation}\label{eq4}
I_{0} - I=I_{0}(1 - e^{-\sigma \cdot n \cdot l})
\end{equation}
where $n$ is the number density (in cm$^{-3}$) of H$_{2}$O molecules in the sample and $l$ the optical path length, which is  twice the sample thickness because the light is reflected at the sample support.

The number density of H$_{2}$O can be computed from the density $d$ of  amorphous ice (0.94~g~cm$^{-3}$)  \cite{Murray2000} taking into account that a 1 to 10 mixture of CO and H$_{2}$O contains 91$\%$ H$_{2}$O, so that $n$=$N_{\rm{av}}$ $\cdot$ $d$ $\cdot$ 0.91/m$_{\rm{H_{2}O}}$, where m$_{\rm{H_{2}O}}$ is the molar mass of H$_{2}$O.
Noting that the mixing ratio used is 10, we make the approximation that any H$_{2}$O molecule in the experimental sample that is photodissociated will be close to one CO molecule with which it can react, as in our model. Errors due to this approximation are not likely to exceed a factor 2.

In Watanabe's experiment \cite{Watanabe2002}, the measured photon flux is $I\rm{_{0}}$=9.2$\times$10$^{13}$ photons~s$^{-1}$~cm$^{-2}$, and the estimated double path lenght is $l$=1.6$\times$10$^{-6}$~cm. Thus, applying Eq.~\ref{eq4}   the probability that a UV photon will be absorbed in the experiments  should  be about 0.28, and per second the number of OH radicals formed in the experimental ice sample ($l$=1.6$\times$10$^{-6}$~cm)  per surface area of 1~cm$^{2}$   should be   2.3$\times$10$^{13}$ ($N_{\rm{OH}}$), which may react with CO molecules. 
The number of CO molecules ($N_{\rm{CO}}$) in a sample of 0.8$\times$10$^{-6}$~cm  is equal to \cite{Watanabe2002}  $N_{\rm{CO}}$=0.1 $\cdot$ $n$ $\cdot$ 0.8$\times$10$^{-6}$ = 2.32$\times$10$^{15}$.

Watanabe $et~al.$ \cite{Watanabe2002}  measured   the time variation  of the  integrated absorbance  of CO molecules with respect to the initial amount of CO molecules (see figure 3 in Ref. \cite{Watanabe2002}) and also the time variation  of integrated absorbance of CO$_{2}$ normalized to the initial integrated absorbance of CO (see figure 4 in Ref. \cite{Watanabe2002} ). From the fits of the  two curve (figures 3 and 4 in Ref. \cite{Watanabe2002}) they obtained  rate constants equal to 6.5$\times$10$^{-4}$~s$^{-1}$, and 2.5$\times$10$^{-4}$~s$^{-1}$ for the time  destruction of CO and formation of CO$_{2}$, respectively. 
From the CO destruction rate 6.5$\times$10$^{-4}$~s$^{-1}$, we derive that per second 1.5$\times$10$^{12}$ CO molecules  are converted, either to form HOCO or CO$_{2}$.
Summarizing, in the experiment for each OH formed, 1.5$\times$10$^{12}$/2.3$\times$10$^{13}$= 0.07 molecule CO disappears. This number is significantly larger than the value obtained in the simulations    (3.6 $\pm$ 0.7)$\times$10$^{-4}$.
On the other hand, the experimentalists  concluded that if CO converts immediately to CO$_{2}$ both rate constants should be the same, but they are not. 
 Therefore, in their experiments  there must be some intermediate in the solid state before CO$_{2}$ is formed. This intermediate could be the HOCO complex even though it was not observed in their FTIR measurements.

In our simulations, the probability to destroy CO and form  CO$_{2}$ is  much lower   than the experimental value, if we assume that in the experiments all destroyed CO molecules correspond to CO$_{2}$ formation (see Table~\ref{table1}).
However, our results suggest that the HOCO complex is formed in the ice and that it is   stabilized by energy dissipation.
Therefore, if we compare the total calculated probability (Table~\ref{table1})  to form HOCO and CO$_{2}$ with the experimental value assuming that all the CO that is destroyed  results into HOCO or CO$_{2}$ the agreement is rather good, even though there was no experimental evidence for large  HOCO concentrations inside the ice in this particular experiment.
But in a later experiment Watanabe $et~al.$ \cite{Watanabe2007} found two peaks at 1808 and 1784 cm$^{-1}$ that they decided not to investigate. However, we believe that these two peaks correspond to the $trans$-HOCO and $cis$-HOCO complex with frequencies of 1774 cm$^{-1}$ and 1812 cm$^{-1}$, respectively according to Oba $et~al.$ \cite{Oba2010}, and 1797 cm$^{-1}$, and 1833 cm $^{-1}$ (measured in a matrix), respectively,  according to Milligan \& Jacox \cite{Milligan1971}. Thus, it is possible that these features were also present  in the earlier experiments of Watanabe $et~al.$  \cite{Watanabe2002} but not reported.

%

\begin{table}[b]
\caption{Experimental \cite{Watanabe2002} and theoretical probabilities after CO and OH reaction in the ice.}
\begin{center}
\begin{tabular} { c   || c | c  } 
\hline
\hline
Channel & Experim. probability \cite{Watanabe2002}  & Theoret. probability (this work) \\ \hline
Total CO conversion  & 7.00$\times$10$^{-2}$  & (3.01 $\pm$ 0.07)$\times$10$^{-2}$    \\
CO$_{2}$ and H formation   &  2.69$\times$10$^{-2}$  & (3.6 $\pm$ 0.7)$\times$10$^{-4}$    \\  
HOCO formation   & - &   (3.0 $\pm$ 0.1)$\times$10$^{-2}$    \\ \hline
\end{tabular}
\end{center} 

\label{table1}
\end{table}

A large difference between the theory and the experiments is that in the theory only one CO molecule and one nearby H$_{2}$O molecule excited by a single photon are considered, whereas in the laboratory conditions several OH radicals are produced that can
interact with several CO molecules, and find the ideal orientation and react. Another important difference is that in our theoretical study we  excite H$_{2}$O molecules to the first excited state ($\rm{\tilde{A}^{1}B_{1}}$) leading to H and OH, with the OH radical  in its electronic  ground state ($X$). 
However, in the experiments  the ice is irradiated  with photons with energies close to the Ly$\alpha$ energy,  which  includes absorption into the $\rm{\tilde{B}^{1}A_{1}}$ state and may lead to the formation of: OH(X) + H, OH(A) + H, O($^{3}$P) + 2H,  and O($^{1}$D) + H$_{2}$, with the corresponding experimental branching ratios of: 0.64, 0.14, 0.22, and 0, respectively,  according to Mordaunt $et~al.$ \cite{waterBstate}. 
It is also known from gas-phase experiments \cite{Harrich2000}  and theoretical calculations \cite{Fillion2001} that the  OH(X) product has  a vibrational state distribution  mostly only populated at $\upsilon$=0, whereas the OH(A) product after H$_{2}$O photodissociation at Ly$\alpha$ is formed not only in  $\upsilon$=0, but also in $\upsilon$=1 and 2 in the gas-phase.
Since the OH(X) radicals in  $\upsilon$=0 are not so reactive (Fig.~\ref{Figure5}), the OH radicals  in the  excited state ($A$) 
may well be   responsible for the reaction with CO in the experiments. 

\subsection{Improvements of the model} 
\label{ssec:Improve}
In our previous studies \cite{Andersson2006, Andersson2008} we already discussed the most important approximations introduced in the study of molecular dynamics simulations of water ice photodissociation. Here, we discuss the ones that may effect our results regarding the formation of CO$_{2}$ through the reaction of CO with OH formed upon H$_{2}$O photodissociation. We list them as follows:
\begin{enumerate}

\item The formation of CO$_{2}$ ice has been studied only through the CO + OH(X) route.  
The photoexcitation of H$_{2}$O  to the first electronic excited state only leads to OH(X) molecule in its ground state and H atom.
In the experiments  H$_{2}$O can also be  photoexcited to the second excited electronic state and  its dissociation can lead to the formation of OH(X) + H, OH(A) + H, O($^{3}$P) + 2H, and O($^{1}$D) + H. In principle, two additional routes may then lead to the formation of CO$_{2}$: (1) CO + O($^{3}$P) $\rightarrow$ CO$_{2}$, which can also take place when the O($^{3}$P) is formed after the reaction of two OH radicals formed upon independent H$_{2}$O dissociation processes, (2) CO + OH(A) $\rightarrow$  CO$_{2}$ + H, which probably has a higher associated reaction probability than the CO + OH(X) reaction, as we already discussed in the previous section. 
Therefore, a more precise comparison with the experiments could be enabled by also modeling the excitation of H$_{2}$O to the $\rm{\tilde{B}}$ state in our simulations. Also, experiments could be carried out  in which H$_{2}$O ice is irradiated at lower excitation energies, so that only excitation to the  $\rm{\tilde{A}}$  state is possible, which would enable a better comparison to the present theoretical work. 
Another difference with the experiments is the CO/H$_{2}$O ratio being equal to 10  which is much larger than in our model. Therefore, in the experimental ice sample, the H$_{2}$O molecules can probably rotate rotate more easily and eventually find the best orientation to react with CO.
On the other hand, our model is well suited to the description of interstellar chemistry in terms of photon fluxes (which are very low, meaning that the photodissociation by one incident photon is completely finished by the time the next photon arrives). But with the interstellar radiation field, excitation should also be possible to the  $\rm{\tilde{B}}$  state of H$_{2}$O.

\item The particular  HOCO PES used in the model may  introduce some errors in the reactivity. The HOCO PES that we considered is the LTSH PES \cite{LTSH}.
Gas-phase calculations \cite{Li2012b} computed for the HO + CO reaction by means of different PESs showed differences in the cross sections obtained, but only at collision energies E$_{c}$ $>$ 0.65 eV (15 kcal/mol). However, for the collision energies we are considering (E$_{c}$ $<$ 0.17 eV (3.9 kcal/mol)) the differences in the cross sections obtained with different PESs are almost negligible (see Fig.4(a) in Ref.~\cite{Li2012b}).

\item Not including quantum effects may be important at the low temperatures here considered. However, in the gas-phase calculations \cite{Li2012b} quantum effects are small, therefore we assume they will also be  small in the solid state environment  considered here.

\item  In our model we have treated the interaction between H and CO through a non reactive HCO PES. Thus, no competition with HCO formation has been introduced, and in the future we are planning to also include a reactive HCO PES in order to quantify the probability of HCO ice formation in the interstellar medium. It should be noted that HCO formation would probably occur at the cost of the CO$_{2}$ formation.

\end{enumerate} 


\section{Summary and Conclusions} 
\label{sec:Conclusions}

Molecular dynamics simulations have been performed for two different  CO--H$_{2}$O ice systems in order to study the formation of CO$_{2}$ in interstellar ices upon  irradiation with  UV photons. After the absorption of an UV photon H$_{2}$O dissociates into H and OH, and the OH radical can react with the CO present in the ice and form CO$_{2}$.

The first type of ice is based on the `hit and stick'  method with a single CO molecule in the center and 50 water molecules around it yielding  an ice ball.
Three different initial ice balls were set up at 10~K and several H$_{2}$O molecules were chosen to be photodissociated. We calculated the probabilities to form the HOCO complex and the CO$_{2}$ molecule, and also the probability to form the  transient  HCO van der Waals complex.

The  second kind of ice was based on our previous MD simulations of amorphous water ice dissociation \cite{Andersson2006, Andersson2008, Arasa2010}, and by using the same procedure an amorphous water ice at 10~K was set up with either a CO molecule  adsorbed on the topmost ML or  located  in the second or fifth monolayer. Several initial conditions were sampled for the three cases, and no reaction was observed in the first case (CO$_{\rm{ad}}$ adsorbed on the surface). In the second case (CO$_{\rm{ab}}$ in ML~2) only the formation of the HCO van der Waals complex was registered  (even though a non reactive HCO PES was used), and in the end this resulted in the H atom desorbing from the ice and the CO molecule being trapped inside the ice. However, if  the CO was initially absorbed in ML~5, the HOCO complex and the CO$_{2}$ molecule can be formed and the probabilities of these events were calculated.

The formation  of the HOCO complex is much more probable than the formation of CO$_{2}$, because once the HOCO complex is formed it immediately starts  losing its internal energy and it may become trapped inside the ice matrix. The HOCO complex has  been detected experimentally by Oba $et~al.$ \cite{Oba2010}, and Watanabe $et~al.$ \cite{Watanabe2002} also claimed that this complex must be formed   as an intermediate that in the end may dissociate into  CO$_{2}$ and hydrogen atom. 
The comparison with the experiments is not straightforward mostly because the experimentalists irradiate the ice with a broadlamp peaking near   Ly$\alpha$,  exciting the water molecules to the higher excited $\rm{\tilde{B}}$ state. We have deduced from their data the number of photons that lead to conversion of  CO  into HOCO or further to CO$_{2}$ after irradiating the CO--H$_{2}$O mixed ice with UV photons close to Ly$\alpha$ energies. 
If we assume that all the CO is converted  into CO$_{2}$ in the experiments, the experimental probability is much higher than our probability to form CO$_{2}$. However, if we assume that  CO can be converted to HOCO and/or CO$_{2}$,  our total probability to form both molecules agrees reasonably well with their value.

Overall, we can summarize the conclusions as follows:
\begin{enumerate}
\item The initial location of the CO molecule in the ice surface is very important as found from both
 types of ice simulated. 
This means that not only the environment (e.g., the number of nearest neighbor  water molecules and their  orientations relative to CO),  but also the initial position of the  CO molecule plays a big role. The CO must be located deep in the ice system. In conclusion,  it is the initial position of the CO molecule and not the type of ice we modeled that  appears to affect the  reactivity  of the OH photofragment   with CO.

\item CO$_{2}$ ice can be formed through the OH~+~CO reaction where the OH comes from H$_{2}$O photodissociation, if the water molecule has been excited with energies higher than 8.6~eV. However, the HOCO complex is the main product in our simulations.

\item HCO ice can probably also be formed through the H~+~CO reaction where the H comes from H$_{2}$O photodissociation.

\item Experiments regarding the OH~+~CO reaction upon  excitation of H$_{2}$O in the first absorption band of amorphous water ice are necessary in order to confirm  our theoretical results. If H$_{2}$O is excited in the second absorption band (as has been done in all existing experiments) other channels besides the OH(X) + H are possible after H$_{2}$O dissociation, and even  the OH(X) formed will likely  have a different reactivity with CO, due to a higher initial vibrational energy content.

\item A new theoretical model able to photoexcite H$_{2}$O to be $\rm{\tilde{B}}$ state is needed in order to explore other possible routes to form CO$_{2}$ ice.

\end{enumerate}

\acknowledgement
The authors would like to thank S. Andersson for providing the $ab~initio$ data and the fitting of some of the potential energy surfaces employed in our model, and R. Valero for providing the implemented LTSH PES in the VENUS code. This project was funded with  computer time by NCF/NWO, and by  NWO astrochemistry grant No. 648.000.010.

\suppinfo

The parameters of the pair potentials and its analytical expressions,  the analytical expressions of the switching funcions,
and the operational definition of the outcome products (5 pages).

\newpage

\newpage

\begin{figure}[t]
\begin{center}
\includegraphics[width=10cm]{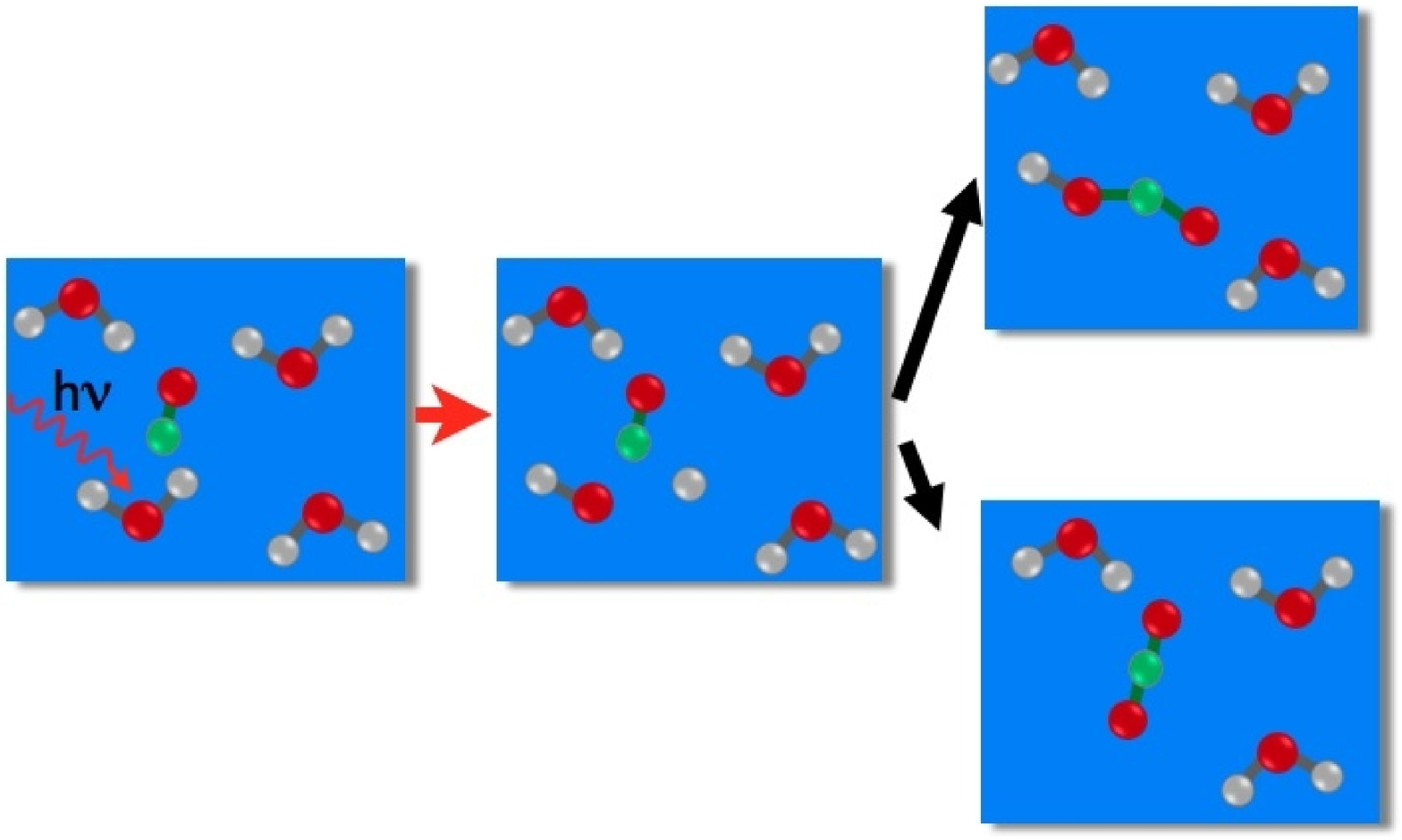}
\end{center}
\end{figure}  

\end{abstract}

\end{document}